\documentclass[seceq]{ptptex}

\title{
From Sakata Model to Goldberg-Ne'eman Quarks and Nambu QCD  
\\
Phenomenology and ``Right" and ``Wrong" experiments%
}

\author{
Harry J. \textsc{LIPKIN}%
}

\inst{
Department of Particle Physics \\
Weizmann Institute of Science, Rehovot 76100, Israel\\
 and\\
School of Physics and Astronomy \\
Raymond and Beverly Sackler Faculty of Exact Sciences \\
Tel Aviv University, Tel Aviv, Israel\\
and\\
High Energy Physics Division, Argonne National Laboratory \\
Argonne, IL 60439-4815, USA
}

\abst{

The basic theoretical milestones were the Sakata SU(3) symmetry, the
Goldberg-Ne'eman composite model with SU(3) triplets having baryon number (1/3)
and the Nambu color gauge Lagrangian.  The transition was led in right and
wrong directions by experiments interpreted by phenomenology. A ``good"
experiment on $\bar p p$ annihilation at rest showed that the Sakata model
predictions disagreed with experiment. A ``bad" experiment prevented the use of
the Goldberg-Ne'eman triplet model to predict the existence and masses of the of
the $\Xi^*$ and $\Omega^-$. More ``good" experiments revealed the existence and
mass of the $\Xi^*$ and the $\Omega^-$ and the absence of positive strangeness
baryon resonances, thus confirming the ``tenfold way". Further ``good
experiments" revealed the existence of the vector meson nonet, SU(3) breaking
with singlet-octet mixing  and the suppression of the $\phi \rightarrow \rho
\pi$ decay.   These led to the quark triplet model. The paradox of peculiar
statistics then arose as  the $\Delta^{++}$ and $\Omega^-$ contained three
identical spin-1/2 fermions coupled symmetrically to spin (3/2). This led to
color and the Nambu QCD.     The book ``Lie Groups for Pedestrians"
used the Sakata model with the name  ``sakaton"  for the $pn\Lambda$ triplet to
teach the algebra of SU(3) to particle physicists in the U.S. and Europe who
knew no group theory.      The Sakata model had a renaissance in hypernuclear
physics in the 1970's.
}

\begin{document}

\maketitle

\section {The Role of experiments, right and wrong}

Physics is an experimental science. The models and ideas that remain are
determined by  experimental tests. Right experiments can disprove wrong models.
Wrong experiments can  lead theorists astray. Both of these occurred in the
transition from the Sakata model\cite{sakata,ikeda,sawada} to the 
Goldberg-Ne'eman-Gell-Mann-Zweig quark model\cite{GOLDNEEM}.

The same SU(3) ``unitary symmetry" group was used both in the Sakata model and 
in the Gell-Mann-Ne'eman octet model called the ``Eightfold
Way".

$SU(3)$ is a natural symmetry of the Sakata model which is built on a
fundamental  triplet. But there is no obvious fundamental triplet in the
eightfold way.
Goldberg and Ne'eman showed how to build the baryon octet from
fundamental triplets\cite{GOLDNEEM}.

A ``Right Experiment" \cite{LevSal} showed that the Sakata model disagreed with 
experiment.
The reaction $ \bar p p \rightarrow K_L K_S $ forbidden by Sakata Model was found.
 
A ``Wrong Experiment" showed that Goldberg-Ne'eman triplets  disagreed  with
experiment.
The decay $ \Sigma^* \rightarrow \Sigma \pi $ allowed by G-N was 
not found\cite{sakurai}. 
\begin{equation}
\frac{BR [\Sigma(1385) \rightarrow \Sigma\pi]}{
BR [\Sigma(1385) \rightarrow \Lambda\pi]} = (4 \pm 4) \% 
\label{wrongexp}
\end{equation}
This led to a wrong selection rule\cite{sakurai} forbidding 
$ \Sigma^* \rightarrow
\Sigma \pi $ and requiring the
$\Sigma^*$ to be in an exotic 27 - dimensional representation of SU(3). 

\section{From Sakata to Goldberg-Ne'eman, Gell-Mann, Zweig, Nambu}

\subsection{Experiments and phenomenology}

The impact of the Sakata model in the period before 1964 has been described by 
Okun\cite{okun}. This paper considers the period beginning in 1961, describes
the impact of the Sakata model in the overlapping period 1961-64 from a very
different perspective and continues on beyond. 

The steps leading from the Sakata model to the quark  model via the
Goldberg-Ne'eman model and eventually to Nambu QCD were led by experiments
interpreted by phenomenology.

The Goldberg-Ne'eman paper $anticipated$ the quark model. All the basic physics
of constructing the baryon octet from three triplets with baryon number (1/3)
is in the paper. That the triplets must have fractional charge if they are
identical is obvious to any graduate student who understands group theory. 
That the U(3) classification of mesons required nonets containing a singlet as
well as an octet of SU(3) should also have been obvious. One might also have
seen that the $K-\pi$ mass difference required breaking  $U(3)$ to $U(2)\otimes
U(1)$ and what was later was called ``ideal mixing" of  the SU(3) singlet and
octet vector mesons. But the paper was a bit ahead of its time. The
experimental situation was not ready for its acceptance. 

The wrong experiment (\ref{wrongexp}) pushed theorists into classifying the  
$\Delta (1238)$ and   $\Sigma(1385)$ in the 27-plet of SU(3). This pushed
experimenters to search for their partners in the 27-plet, the $K^+N$
resonances which were  not found.

The discovery of the $\Xi^*$ (1530) led  Glashow and Sakurai\cite{glashsak} to
classify it together with the $\Delta (1238)$ and   $\Sigma(1385)$ in ``The
Tenfold Way" and predict the existence of an isoscalar baryon with strangeness
(-3) and a mass low enough to make it stable against strong decays. This was
also noted by some others, including Gell-Mann and Ne'eman, who pointed it out
informally at the ICHEP (Rochester) conference in Geneva in 1962. But the first
published calculation was by Glashow and Sakurai who also did a detailed
analysis with an estimate of its possible production in cosmic rays, and found a
serious candidate event. To an unbiased observer, the credit for the prediction
of the existence of the $Z^-$, now called the $\Omega^-$, belongs to Glashow and
Sakurai.

Another crucial experimental development was the discovery of the $\phi$ meson
whose decay $\phi\rightarrow K\bar K$ was SU(3) forbidden for a unitary singlet.
The $\phi$ had to be a mixture of an SU(3) singlet and octet produced by SU(3)
breaking. There was also
a peculiar suppression of the SU(3)-allowed $\phi\rightarrow \rho \pi$ decay.

A simple dynamical model\cite{katzlip} used the $K-\pi$ mass difference to break
SU(3), mix the SU(3) singlet and octet vector mesons $\omega_1$ and $\omega_8$
and forbid the $\phi\rightarrow \rho \pi$ decay. The loop diagram connecting
$\omega_1$ and $\omega_8$ vector mesons via vector-pseudoscalar intermediate
states would cancel in the SU(3) symmetry limit   
\begin{equation}
\omega_1 \rightarrow \rho \pi \rightarrow \omega_8; ~ ~ ~  
\omega_1 \rightarrow K^* \bar K \rightarrow \omega_8 ~ ~ ~  
\omega_1 \rightarrow \omega_8 \eta_8 \rightarrow \omega_8
\label{vecmix}
\end{equation}
However, the SU(3) breaking which lowers the pion mass far below the $K$ and
$\eta$ enhances the $\rho-\pi$ propagator
in the transitions (\ref{vecmix}). The $2 \times 2$ matrix (\ref{vecmix})
is thus dominated
by transitions via the $\rho-\pi$ intermediate state. 
Diagonalization of the loop-diagram  matrix (\ref{vecmix})
with SU(3) breaking expressed by inserting experimental masses
in the propagators produced mixed $\omega_1-\omega_8$ eigenstates with one 
eigenstate, the $\phi$, approximately decoupled from the $\rho-\pi$ channel.

The U(3) description was rediscovered by Okubo
who had noted that enlarging SU(3) symmetry to U(3) produced a meson nonet 
and that breaking 
the $U(3)$ to $U(2)\otimes
U(1)$ gave what was later was called ``ideal mixing"
\cite{Okubophi} of  the SU(3) singlet and
octet vector mesons and naturally suppressed the
$\phi\rightarrow \rho \pi$ decay. 

\subsection{Quarks and Aces}

These new experimental developments set the stage for a phenomenological
investigation of the basic physics of the Goldberg-Ne'man model in which hadrons
were constructed from a fundamental U(3) triplet with baryon number (1/3). The
new phenomenology, called ``quarks" by Gell-Mann\cite{MGM} and 
``aces" by Zweig\cite{Zweig} showed remarkable agreement with the new
experiments. The ``Goldhaber Gap" in the experimental data showed the
absence of the $K^+N$ resonances and ruled out the 27-plet. The baryon spectrum
confirmed the ``tenfold way". The
vector nonet was found and the $\phi-\omega$ mixing was observed and understood.

The statistics problem remained open. The $\Delta^{++}$, $\Delta^{-}$ and
$\Omega^-$ apparently  violated fermi statistics by containing three identical
spin-1/2 fermions in a spatially symmetric S-state coupled  symmetrically to
spin (3/2). This was solved by Greenberg\cite{OWG} by the introduction of
parastatistics which later was seen to be equivalent to the introduction of
another degree of freedom, now called color.

But the new phenomenology still had no sound theoretical basis. And there was no
explanation for the peculiar hadron spectrum which had only
quark-antiquark mesons and three-quark baryons and no ``exotic states" with more
quarks and antiquarks. Since both $qq$ and $q \bar q$
interactions were attractive, there seemed to be no simple way to prevent 
an antiquark from being bound to the three quarks in a baryon by the strong
$q\bar q$ attractive force to make a $3q\bar q$ hadron. 

\subsection{The road to QCD}

The new theory was supplied by Nambu\cite{Nambu}, who showed that all strong
interaction physics could be described by a non-abelian color-gauge theory with
a Lagrangian now called the Lagrangian of QCD. 

Nambu's QCD thus solved the three basic yet unsolved puzzles in strong
interaction physics\cite{triex}:
\begin{enumerate}
\item[1)] The triality puzzle. Why only states of triality zero appear as bound
hadronic states.
\item[2)] The meson-baryon puzzle. Why the $qq$ and $q \bar q$ interactions are
both attractive but differ in strength in a way to bind the two-body $q \bar q$ 
system and the three-body $qqq$ system.
\item[3)] The exotics puzzle. Why only the $q \bar q$ and $qqq$ systems have
bound states and all hadrons  containing more quarks would decay into mesons
and baryons 
\end{enumerate}

\section{The crucial experiments and their phenomenological interpretations}

\subsection{The right experiment that killed the Sakata Model}

During the  winter of 1961-62 The Weizmann Institute group  was investigating
experimental tests of unitary  symmetry to distinguish between the Sakata and
octet models\cite{LLMsakoctet}. At a small  conference on  unitary symmetry 
organized by  Abdus Salam  at Imperial College, Harry Lipkin reported on the
application to $\bar p p$ annihilation into two mesons.  Calculations on the 
blackboard with Salam showed that new  experimental results  from CERN strongly
favored  the Sakata model.    Lipkin returned to Rehovot and discussions with
Carl Levinson and Sydney Meshkov  immediately revealed that the Salam - Lipkin
calculation was based on incorrect values of SU(3)  Clebsch-Gordan
coefficients. The correct values gave a much more exciting and opposite result;
a strong disagreement between the predictions of the Sakata model and the new
experimental data.

This left a quandary.  The paper had to be written immediately and sent for
publication. Salam's name could not be omitted, since  he had participated in
the discussion that had led to this  work. But including his  name as an author
would make him responsible for conclusions which were the exact opposite of his
understanding from the meeting.  Salam had already left for Pakistan and there
were no postal relations  between Israel and Pakistan. The manuscript was sent
to  Gerry Brown, then starting a new journal called ``Physics Letters", with
an explanation of the  problem and carte blanche  to use  his own  judgement.
Brown contacted people at Imperial  College  and found that the error had
independently  been discovered by  a member  of Salam's group.  The publication
appeared in Volume 1 of Physics Letters with the three from  Weizmann, Salam
and his collaborator as joint authors. 

\subsection{The wrong experiment that led us to miss quarks}

Some time in the academic year 1961-62 Hayim Goldberg told about the work he had 
done with Yuval Ne'eman\cite{GOLDNEEM} showing that the baryon octet could be 
constructed from three 
SU(3) triplets with baryon number (1/3). Whether or not you believe that these 
triplets 
are physical objects, this construction is interesting. The obvious (to us now;
perhaps 
not then) is to note that three triplets could make the decuplet, but not the 
27-plet. 
Placing the known resonances now called the $\Delta (1238)$ and the 
$\Sigma(1385)$ in the ten-dimensional representation of SU(3) 
and using the Gell-Mann-Okubo mass formula to calculate the masses would lead 
naturally to the prediction of the existence of the $\Xi(1530)$ and 
the particle now called the
$\Omega^-$ with masses close to those eventually observed.

However this was not considered because the experimental data, now known to be
wrong,  indicated that the decay $\Sigma(1385) \rightarrow \Sigma\pi$  was
forbidden.  This selection rule forced the classification of the $\Delta
(1238)$ and the  $\Sigma(1385)$ in the 27-dimensional representation of SU(3)
and not in the 10.

Sakurai \cite{sakurai} had  noted that the experimental value (\ref{wrongexp})
implied a selection rule forbidding the $\Sigma\pi$ decay. 
Incorporating a symmetry of hypercharge reflection  called R-invariance into
the ``Eightfold Way" gave this selection rule\cite{sakurai} and required
putting the $\Delta (1238)$ and the  $\Sigma(1385)$ in a 27-plet and not in a
decuplet.

A detailed SU(3) description given by Okubo \cite{okubo} noted that
the  $\Sigma(1385) \rightarrow\Sigma\pi$ decay was forbidden for a 27-plet
$\Sigma(1385)$ and the result for the decuplet strongly disagreed with
experiment.

\begin{equation}
\frac{BR [\Sigma(27) \rightarrow \Sigma\pi]}{
BR [\Sigma(27) \rightarrow \Lambda\pi]} = 0; ~ ~ ~  
\frac{BR [\Sigma(10) \rightarrow \Sigma\pi]}{
BR [\Sigma(10) \rightarrow \Lambda\pi]} = 15 \% 
\end{equation}

The Weizmann group\cite{LLMsakoctet} saw
that the 27-plet
classification was needed to fit experiment and immediately called for 
experimental searches for the positive strangeness resonances
expected in the 27-plet but not in the decuplet. 

Unfortunately these data were wrong, there was no selection rule and no $K^+N$
resonance. The new data much later confirmed the decuplet branching ratio.

\begin{equation}
BR [\Sigma(1385) \rightarrow \Sigma\pi]  = (11.7 \pm 1.5) \%; ~ ~ ~ 
BR [\Sigma(1385) \rightarrow \Lambda\pi] = (87.0 \pm 1.5) \%
\end{equation}
\begin{equation}
\frac{BR [\Sigma(1385) \rightarrow \Sigma\pi]}{
BR [\Sigma(1385) \rightarrow \Lambda\pi]} = (13 \pm  2)\% 
\end{equation}

\subsection{Further ``right experiments" that confirmed the triplet model}

New ``right" experiments found the $\Xi^*$ (1530) and revealed the complete 
absence of the positive strangeness KN resonances expected in the 27-plet,
called the Goldhaber Gap. But their implications for
the decuplet classification were not noted until the $\Xi^*$ was found, rather
than noting that the existence and mass of the $\Xi^*$ should have been
predicted. 
The wrong value (\ref{wrongexp}) for the $\Sigma\pi$ decay prevented seeing
the obvious implications of the Goldberg-Ne'eman breakthrough. 

That the $\Xi^*$ mass fit exactly the prediction of the  Gell-Mann-Okubo mass
formula for a decuplet was immediately noted by Glashow and
Sakurai\cite{glashsak}.
Their ``tenfold way" paper was  immediately noted and used\cite{MLLUspin} to make
SU(3) predictions for decuplet production in meson-baryon reactions, and the
possibility of making $Z^-\bar Z^+$ pairs in nucleon-antinucleon annihilation.
This first published prediction\cite{glashsak}  for the existence of this
particle was then already acknowledged as $Z^-$ in published literature, but is 
now generally overlooked.  I once asked J. J. Sakurai why they
never claimed credit for the first publication of this prediction. His response
was that they were highly embarrassed by their paper because they had blindly
substituted into the Gell-Mann-Okubo mass formula without noting that this
becomes an equal-spacing rule for a decuplet.  

The finding of this particle, now called the $\Omega^-$, together with the 
``Goldhaber Gap" confirmed the
``Tenfold Way" now called the decuplet classification and led to the general
acceptance of the Goldberg-Ne'eman triplet model, now called the quark model.
But the wrong experimental value (\ref{wrongexp}) 
for the $\Sigma\pi$ decay remained an obstacle to this interpretation until
better experiments showed agreement with the decuplet prediction.

\section{The impact of the Sakata Model beyond the original $pn\Lambda$ hadron
Model}

\subsection{The ``sakaton" - teaching group theory to particle physicists}

The SU(3) ``unitary symmetry" group used in the Sakata model was also used  in
the Gell-Mann-Ne'eman octet model called the ``Eightfold Way". But Murray
Gell-Mann, like most particle theorists in the U.S. and Europe, knew no group
theory at the time. Group theory was viewed as irrelevant mathematics (Die
Gruppenpest) which had no use in particle physics. And nuclear and condensed
matter physics were disregarded as ``dirt physics" and ``squalid state 
physics"\cite{Gross}. The particle theorists saw
isospin as rotations in an abstract  three-dimensional space and spent eight
years searching for a higher symmetry in  rotations in spaces of higher
dimensions\cite{Lipgroup}.  One might say that they called SU(3) symmetry the
eightfold way because it took them eight years to learn that  isospin and
strangeness  are $SU(2) \otimes U(1)$. They did not know that  isospin is also
$SU(2)$  and that $SU(3)$ is a natural symmetry to include $SU(2)$ and $U(1)$. 

The existence and algebra of unitary groups and in particular SU(3), although
unknown to Gell-Mann and his American and European colleagues in 
particle physics, was well known in the nuclear physics community. 
However SU(3) was used in
nuclear physics as the invariance group of the three-dimensional harmonic
oscillator, and its representations and Clebsch-Gordan coefficients 
were always classified using the subgroup
$O(3)$ of rotations in three dimensions. Levinson, Lipkin and Meshkov knew this
classification very well, but relied on the Sakata model papers\cite{ikeda,
sawada} to obtain the SU(3) Clebsch-Gordan coefficients using the subgroup
$SU(2)\times U(1)$. They used the word ``sakaton" as a general name for a
fundamental triplet of SU(3).  

The book ``Lie Groups for Pedestrians" \cite{LGP} arose from the need for a 
simple set of lectures to teach the necessary Lie algebras to particle and
nuclear physicists. To present SU(3) in a simple way,  the Sakata model was
used with the name  ``sakaton" for  the $pn\Lambda$ triplet.

\subsection{Renaissance of the Sakata model in hypernuclear physics} 

In 1971 a physical motivation was presented for using a dynamical symmetry of
the Sakata model type in hypernuclear physics was given\cite{Kermlip}. The
$pn\Lambda$ triplet was considered as the constituents of hypernuclei with
SU(3) symmetry. Earlier works on SU(3) symmetry applications to hypernuclei had
used the octet SU(3) version in which the $\Lambda$ and $\Sigma$ are
degenerate. But the 80 MeV $\Lambda - \Sigma$ splitting was too big for nuclear
and hypernuclear excitations.

A ``strangeness analog state"
obtained by a U-spin operation on all neutrons in the nuclear ground state
was defined. The suggestion\cite{schiflip} that this strangeness analog state
had been observed in the first ($K^-,\pi^-$) experiments done in the CERN PS.
turned out to be wrong.  Dalitz and
Gal\cite{dalgal} narrowed the symmetry to only the valence neutrons.
The Sakata SU(3) symmetry was combined with Pauli spin
SU(2) to SU(6) supermultiplets which include both nuclei and hypernuclei,
but only for a particular shell correspondence. This was a natural extension
of Wigner's SU(4) supermultiplet theory for ordinary (light) nuclei.
In particular the $^{9}_{\Lambda}{\rm Be}$ hypernucleus was 
analyzed\cite{dalgal2}.
A very interesting consequence of this work concerned a particularly symmetric
state in the excited hypernuclear spectrum termed ``supersymmetric".
This state has been discovered in $^{9}_{\Lambda}{\rm Be}$. 

This supersymmetric state concept was rediscovered in 1983 and termed 
a ``genuinely hypernuclear state"\cite{japsup}. A review\cite{hypjap}cites both the 
Dalitz-Gal theoretical work\cite{dalgal} in 1976 and the 
Japanese work\cite {japsup} in the 1980s.

\section{Details of the Sakata selection rule forbidding $p \bar p \rightarrow 
K_L K_S$}

  In the Sakata model annihilation into charged kaon pairs and charged pion pairs is 
allowed but annihilation at rest into neutral kaon pairs is forbidden. 
This prediction was  in strong disagreement with experiment,   which showed that
$K_L - K_S $ pairs were produced at  comparable rates with charged kaon and pion
pairs. There is a very simple ``pedestrian" explanation of this selection rule. 

In the Sakata model the neutral kaons are made of neutrons and $ \Lambda $'s and
their antiparticles and contain no protons  nor antiprotons.   The charged pions
and kaons all contain a proton  or antiproton.   Thus a proton-antiproton system
can become two charged  pions or kaons by creating a  single additional neutron-
antineutron or
 $ \Lambda $-anti-$\Lambda $ pair  which combines with the  initial proton and  
antiproton to form the two final mesons.  This cannot occur for the neutral kaon
pair final state.

   The selection rule can also be seen as  an SU(3)  rotation of the isospin and
parity selection rule  forbidding the annihilation of odd-parity  $ \Lambda \bar
\Lambda $ states into two pions.  The $ \Lambda \bar \Lambda $ state has isospin
zero and the isospin zero states of two pions are symmetric under interchange of
the two pions and have even parity.

 \begin{equation}
\bar \Lambda \Lambda \rightarrow \pi^+ + \pi^- {\rm\ 
(forbidden\ for\  odd\  parity)} 
\end{equation}
   
   In  the  Sakata  model  there is  an  SU(3)   symmetry  transformation  which
interchanges protons and  $ \Lambda $'s everywhere.    Under this transformation
the charged pions,  $(p\bar n)$ and $( \bar pn)$ become neutral kaons $ (\Lambda
\bar n)$ and $(\bar \Lambda n)$ and the selection rule becomes

 \begin{equation}
   \bar pp \rightarrow K^o + \bar {K^o}  {\rm\ (forbidden\ for\ odd\ parity)} 
 \end{equation}

Although this selection rule holds only for odd parity states,  the annihilation
into $ K_LK_S $ pairs is forbidden for all partial waves, since even parity $ K \bar
K $  pairs  are allowed only to  decay only into the  $ K_LK_L $ and  $ K_SK_S $
decay modes, but never into $ K_LK_S $.   In the octet model and the quark model
there is no such  selection rule,  as the analogous transformation  on pions and
kaons via interchanging u  and s quarks mixes $ \Lambda  $'s and $\Sigma$'s
rather than $ \Lambda $'s and protons.

   Salam's collaborator,  Munir Ahmed Rashid, appeared on this
paper\cite{LevSal}  as R. A. Munir instead of M. A.  Rashid. This confusion
arises  because Pakistani  Moslem names  are often  words joined  together in 
a phrase with a well-defined  meaning,  rather than a Christian name  and a
family name. Abdus Salam, for example, can mean a servant of peace.   We
conclude with the hope  that these  days we  should  all try to be servants of
peace.

\section*{Acknowledgements}
It is a pleasure to acknowledge discussions with Avraham Gal, Sydney Meshkov
and Lev Okun. This research  was supported in part by the U.S. Department of
Energy, Division of High Energy Physics, Contract DE-AC02-06CH11357

%

\end{document}